\documentclass[aps,twocolumn,nofootinbib,superscriptaddress]{revtex4-2}
\usepackage{dcolumn}
\usepackage{multirow}
\usepackage{amsmath}
\usepackage{tablefootnote}

\usepackage{graphicx,bm}
\usepackage{bm,color}
\usepackage{float}
\usepackage{mathrsfs}

\usepackage[final]{changes}
\newcommand{\ef}{efm$^2$\,}

\newcommand{\bL}{\begin{Large}}
\newcommand{\eL}{\end{Large}}
\newcommand{\be}{\begin{equation}}  
\newcommand{\ee}{\end{equation}}
\newcommand{\ba}{\begin{eqnarray*}}
\newcommand{\ea}{\end{eqnarray*}}

\begin{document}
\title{Shell model analysis of the $\bm{ B(E2,2^+ \rightarrow 0^+)}$'s in the A=70 T=1 triplet}
%
\author{ S.~M.~Lenzi}
\affiliation{Dipartimento di Fisica e Astronomia, Universit\`a degli Studi di Padova, \\and INFN, Sezione di Padova, I-35131 Padova, Italy  }

\author{A.~Poves} 
\affiliation{Departamento de F\'isica Te\'orica  and IFT-UAM/CSIC, \\Universidad
 Aut\'onoma de Madrid, 28049 Madrid, Spain}

\author{A.~O.~Macchiavelli}
\affiliation{Nuclear Science Division, Lawrence Berkeley National Laboratory\\ Berkeley, California 94720, USA}

\begin{abstract}
The $B(E2,2^+ \rightarrow 0^+)$ transition strengths of the T=1 isobaric triplet $^{70}$Kr, $^{70}$Br, $^{70}$Se, recently measured at RIKEN/RIBF, are discussed in terms of state of the art large scale shell model calculations using the JUN45 and JUN45+LNPS plus Coulomb interactions. \replaced{ In this letter}{, we consider alternative explanations to the possible shape change of  $^{70}$Kr with respect to   Our results question the conclusion that the shape of $^{70}$Kr may be different from that of the other members of the isobaric multiplet.WeIn this letter} we argue that, depending on the effective charges used, the calculations are \added{either} in line with the experimental data within statistical uncertainties, or the  anomaly happens in
$^{70}$Br, rather than $^{70}$Kr. In the latter case, we suggest that it can be due to the presence of a hitherto
undetected 1$^+$ T=0 state below the yrast 2$^+$ T=1 state. \added{ Our results do not support a shape change of $^{70}$Kr with respect to  the other members of the isobaric multiplet.}
 \end{abstract}

\maketitle


\noindent
{\sl Introduction.~}
  In the limit of strict isospin symmetry, the matrix elements 
  \mbox{M$_p$(E2)= $\sqrt{B(E2,0^+\rightarrow 2^+)}$} 
  in a T=1 triplet must vary linearly with the T$_z$ of its members. Nuclear structure details
  determine the slope of the line and the absolute value of the M$_p$(E2)'s. Isospin symmetry breaking effects (ISB)
  are known to produce differences in the binding energies and in the excitation energies of the members of an  isospin multiplet, these 
  are dubbed Coulomb Energy Differences (CED), Mirror Energy Differences (MED) and Triplet Energy Differences (TED),
  respectively \cite{Nolen69,TE1,TE2}. Although the Coulomb
  repulsion among the protons is the main source of these effects, it has been shown that additional ISB terms are needed to explain the available experimental data \cite{Zuker02,Bentley07}.
  However, as the values of the MED's and TED's are quite small, one should expect
  that the isospin breaking effects in the M$_p$(E2)'s be even smaller and that
  linearity should be preserved to a large extent.  Notice, however, that sometimes even a small MED can produce quite prominent effects, as 
  it was the case reported in Ref.~\cite{Hoff20} , where the ground state spins of the mirror pair $^{73}$Sr-$^{73}$Br
  were found to be different. In Ref.~\cite{Lenzi20} we have shown that standard ISB effects  suffice to explain the inversion 
 of two close lying  5/2$^-$  and  1/2$^-$ levels (see also \cite{henderson20}).
 
\medskip
\noindent
In a recent experiment carried out at RIKEN/RIBF~\cite{wimmer2021}, the $B(E2, 0^+\rightarrow 2^+)$'s of the T=1 triplet $^{70}$Kr, $^{70}$Br, $^{70}$Se have been measured. The authors of this work argue that these values, and the extracted M$_p$(E2)'s shown in Fig. 1, are inconsistent with isospin symmetry conservation and suggest that
the shape of $^{70}$Kr may be different from that of the other members of the multiplet \deleted{, without providing any explanation of 
its origin.}
In this work we approach the problem in the framework of the Shell Model with configuration interaction \added{and discuss alternative scenarios}.\\

\noindent
{\sl Large Scale Shell Model Calculations.~}
Two valence spaces are adopted. The first includes the orbits 1$p_{3/2}$, 1$p_{1/2}$, 0$f_{5/2}$, and 0$g_{9/2}$ for both protons and neutrons, 
and we use the effective interaction JUN45\footnote[2]{The same valence space and interaction were used in Ref.~\cite{wimmer2021}.}~\cite{jun45}. The second set of calculations is made in an
extended space which includes the 1$d_{5/2}$ orbit,  with the JUN45 interaction supplemented with the necessary
matrix elements from the LNPS \cite{Lenzi10} interaction and we refer to it as JUN45+LNPS. The calculations were performed with the code Antoine
\cite{rmp} and involve dimensions of O(10$^9$), allowing up to 10p-10h excitations from the  $p_{1/2}$ and $f_{5/2}$ orbits to $gd$ shells, whereas
the jumps from the $p_{3/2}$ are restricted to 4p-4h. \\


\noindent
Concerning the effective charges  (see Ref.~\cite{n=z} for a detailed discussion) we have two
choices:  Dufour-Zuker's (DZ) q$_{\pi}$=1.31e and q$_{\nu}$=0.46e , microscopically derived for harmonic oscillator cores  \cite{dufour}, and 
the standard (ST) ones q$_{\pi}$=1.5e and q$_{\nu}$=0.5e . The latter were shown to be adequate for a   $^{56}$Ni core~\cite{n=z}, but for completeness we will present results with both sets.\\

\noindent
The calculations incorporate the Coulomb interaction between the protons obtained with harmonic oscillator wave functions with the appropriate values 
of $\hbar \omega = 45 A^{-1/3} - 25 A^{-2/3}$[MeV]. As the effects are perturbative  we have computed them in the JUN45 case and add the same corrections to the JUN45+LNPS nuclear only ones. The results with JUN45 plus Coulomb and ST effective charges are given in Table~\ref{tab1}.
While the absolute excitation energies  of the   2$^+$'s are predicted by JUN45 about 300~keV too low with respect to the experiment, the  MED= -100 keV and TED=-17 keV, compare well with the experimental values of  -67.0 (7.5) keV and -45.2 (7.5) keV respectively. The quoted experimental uncertainty corresponds to the root-mean-squared deviation  of the  $^{70}$Kr 2$^+$ state energies given in~\cite{nndcx}. \\

\begin{table}[h]
\caption{A=70 triplet shell model results with the JUN45 interaction plus Coulomb.}
\bigskip
\begin{tabular*}
{\linewidth}{@{\extracolsep{\fill}}|c|ccc|}
\hline
Nucleus&  E(2$^+$)  & M$_p$(E2)  &    $\delta$M$_p$(E2)$_{coul}$ \\ 
             &  [MeV]     & [\ef]              & [\ef]  \\  
\hline 

         $^{70}$Kr  &  0.545 & 52.2 &  +3.4 \\
         $^{70}$Br & 0.605  &  47.0 &  +0.7 \\
         $^{70}$Se& 0.648 & 43.5 & -0.3 \\
   \hline
\end{tabular*}
\label{tab1}
\end{table}   

\begin{figure}[h]
\begin{center}
\includegraphics[width=0.8\columnwidth]{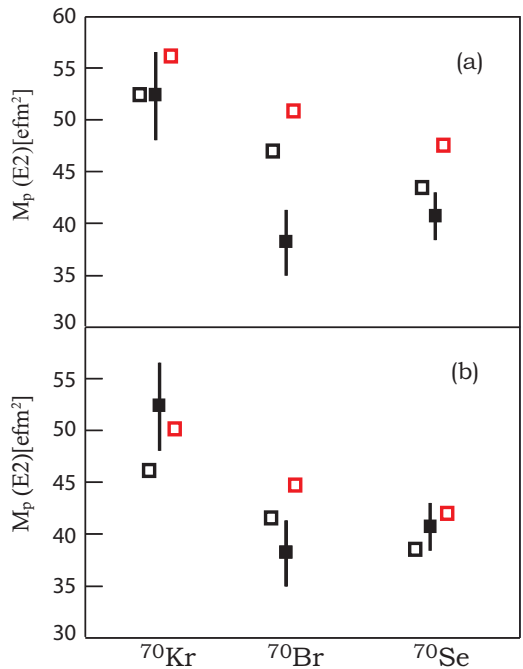}
\caption{M$_p$(E2) in the A=70 isospin triplet (in efm$^2$). The experimental data are taken from Ref. \cite{wimmer2021} (See text for a discussion of the
error bars). Results of the JUN45 (black squares) and JUN45+LNPS (red squares) calculations. (a) ST effective charges; (b) DZ effective charges. 
\label{me2}}
\end{center}
\end{figure}

\noindent
We report in Fig.~\ref{me2} the calculated M$_p$(E2)'s for JUN45 and JUN45+LNPS with both sets of effective charges. The error bars of the experimental points are
taken from Ref. \cite{wimmer2021} for  $^{70}$Kr and are the average of the results of references \cite{wimmer2021,nichols2014,ljungvall2008} for
$^{70}$Br and $^{70}$Se. \\

\noindent
The M$_p$(E2)'s obtained with JUN45+LNPS are very similar to those of JUN45, with the added bonus that it produces a better spectroscopy, with the 2$^+$ energies  closer to the experimental values.
  We see that the Coulomb corrections to the M$_p$(E2)'s are small, but not negligible as
  claimed in \cite{wimmer2021}, mainly 
in the case of $^{70}$Kr. The induced non-linearity goes in the direction demanded by the experimental data.

\medskip
\noindent

     



\noindent

\noindent
We now examine in more detail what Fig.~\ref{me2} tells us. First, let us emphasize that, according to our calculations, none
of the members of the triplet can be said to have a well defined shape. The more so in view of the values of the intrinsic
shape parameters $\beta$ and $\gamma$ and their variances that we obtain from the Kumar invariants \cite{kumar} as described in Ref.~\cite{poves2020}.
The value of $\beta$=0.22$\pm$0.05, means that the fluctuations of Q$^2$ amount to one-half of its mean value. While the value of $\gamma$=32$^{\circ}$ is suggestive
of triaxiallity, its fluctuations at 1$\sigma$ level span the interval 8$^{\circ}$--60$^{\circ}$. Therefore, the certain degree of quadrupole collectivity they exhibit can be accounted without resorting to any shape change, \replaced{in contrast to}{ and} the conclusions in Ref.~\cite{wimmer2021} \deleted{should be questioned}. Second, if we examine the 
lower panel (b) of Fig.~\ref{me2} we realize that the calculations, although marginally, are  compatible with the data. If we consider the upper panel (a), the anomaly, 
if it would exist, could be due to a dip in the M$_p$(E2) of $^{70}$Br. \\

\noindent
We will discuss in the remaining part of the letter a possible explanation 
for this apparent behavior.\\

\begin{figure}
 \begin{center}
\includegraphics[trim=120 190 40 180, clip,width=1.2\columnwidth, angle=0]{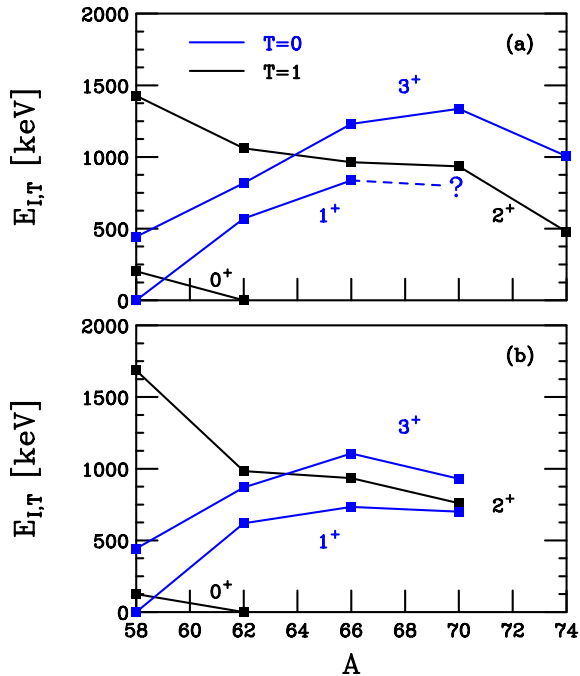}
  \end{center}
\caption{ (a) Systematic of the low-lying T=0 and T=1 states in odd-odd nuclei  relevant to the A=70 region. \\
The 1$^+$ in $^{70}$Br, indicated by a question mark, is not known. (b) JUN45+LNPS  shell model results.
\label{1plus}}
\end{figure}

\noindent
{\sl The structure of $^{70}$Br. ~}
 Despite $^{70}$Br being an N=Z odd-odd nucleus, due to the
role played in its structure by the 0$f_{5/2}$ and 0$g_{9/2}$ orbits, isovector pairing dominates over the symmetry energy and its ground state is 0$^+$ T=1~\cite{aom}. The yrast
2$^+$ T=1 is located at 934~keV (notice that in well deformed $^{76}$Sr it appears at about 200 keV) and the lowest T=0 state, known to date, is a 3$^+$ at 1336~keV~\cite{deAngelis2001,jenkins2002,ljungvall2008,nichols2014,NNDC}. As shown in the systematics in Fig.~\ref{1plus} (a), in other odd-odd N=Z nuclei in the region, the 1$^+$ T=0 state lies a few hundreds of keV  below the 3$^+$ and also below the 2$^+$.
In Fig.~\ref{1plus}(b) we show the results of JUN45+LNPS interaction (spectroscopically superior to those of JUN45) which  remarkably reproduce the experimental trends. \\

\noindent
{\sl The 1$^+$ scenario. ~} While an inspection of  Fig.~\ref{1plus} shows that we should not expect the shell model to predict the excitation energies within  $\approx \pm$200~keV, both systematics and theory suggest that it is possible that the 1$^+$ is lower than the 2$^+$.  In this regard, it is worth  noting that the particle plus rotor model and IBM4 calculations discussed in Ref.~\cite{jenkins2002} also predict the 1$^+$ state lower than the 2$^+$.


\medskip
\noindent
 The key ingredient of this scenario is the $B(M1)$ transition
probability from the  2$^+$  -- T=1 state to the 1$^+$ --  T=0
state, which according to our calculation amounts to   0.22 (2) ~$\mu_N^2$\footnote[3]{These results are obtained with no $g_s$ quenching. For reference, the
theoretical value for the  $B(M1, 3^+ \rightarrow 2^+ )$=0.013~$\mu_N^2$, is compatible
with the experimental value 0.027(12)~$\mu_N^2$ .}.  Thus,  if the experiments did not have enough sensitivity and the M1 transition was not observed,  the $B(E2)$ value, extracted from the intensity of the de-excitation of the 2$^+$ $\gamma$-ray, would have been underestimated.\\

\begin{figure}[h]
\begin{center}
\includegraphics[trim=40 200 20 200, clip,width=1.0\columnwidth,angle=0]{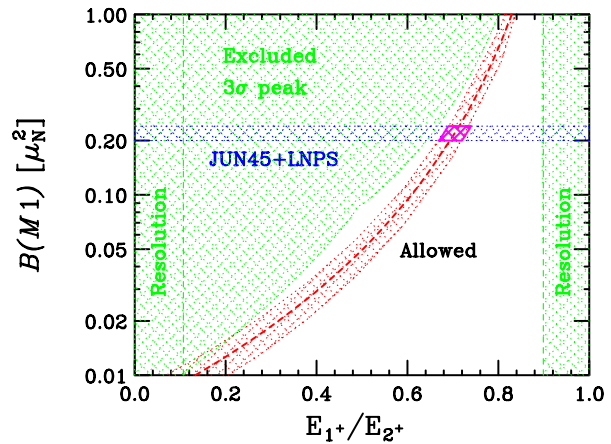} 
\caption{ $B(M1)$ strength from the 2$^+$ to an hypothetical $1^+$ required to explain a missing intensity of the 2$^+$ $\gamma$-ray (red dash line and shaded area) and the excluded regions imposed by the experimental conditions of the setup in Ref.~\cite{wimmer2021} (Green shaded area.  See text for details).   The intersection of the shell model results (blue shaded area) with the empirically required strength determine a possible solution (magenta shaded area).
\label{exclusion}}
\end{center}
\end{figure}

\noindent
To further explore our conjecture, we assume that $^{70}$Br follows isospin symmetry and determine a value of its M$_p$(E2)=46.5 (4.6) \ef from a linear fit of the $^{70}$Kr and $^{70}$Se, $\vert$T$_z\rvert$=1 pair.  For a given energy of the hitherto unknown 
1$^+$ state,  E$_{1^+}$, we then calculate the required $B(M1)$ strength for the transition $2^+\rightarrow 1^+$ such as the
$M1/E2$ branching ratio explains the missing $\gamma$-ray intensity required to obtain the measured M$_p$(E2)=38.1 (3.1) \ef matrix element~\cite{wimmer2021}. The result is shown in Fig.~\ref{exclusion} by the red dash line and shaded area. \\

\noindent
We now consider the conditions of the RIKEN/RIBF experiment, in terms of statistics, peak-to-background and energy resolution to establish the values excluded by the measurements for the observation of a 3$\sigma$ peak in the spectrum. Here, we assume the detection of the higher energy $\gamma$ transition, namely the 2$^+ \rightarrow 1^+$ or the 1$^+ \rightarrow 0^+$.   The results are presented in the form of an exclusion plot (shaded green area) in Fig.~\ref{exclusion}.  Our $B(M1)$ estimates above and the shell model results lie within the allowed region
and a consistent solution exists,   indicated by the intersection of the empirical and shell model values.  \\

\noindent
Furthermore, with the limited information we can assess from Ref.~\cite{jenkins2002}, the statistics of the relevant coincidence spectra shown in the paper seems  consistent with the non-observation of a $\gamma$-ray peak with the intensities allowed by the exclusion plot. Last but not least,  the lifetime measurements of Ref.~\cite{nichols2014} require some discussion.   In contrast to the even-even cases, the line-shape analysis in the odd-odd $^{70}$Br could potentially be more susceptible to the unknown feeding of the 2$^+$ state from T=0 states above.  Therefore, side-feeding corrections, not  fully captured in a singles spectrum, could make the effective lifetime of the 2$^+$ level to appear longer.

\noindent
{\sl Conclusions.~} In summary,  we present  a Large Scale Shell Model analysis of the $B(E2, 2^+ \rightarrow 0^+)$'s in the A=70 T=1 triplet. The calculations were performed using the JUN45 (+LNPS) interactions in the model spaces 1$p_{3/2}$, 1$p_{1/2}$, 0$f_{5/2}$, and 0$g_{9/2}$ (+ 1$d_{5/2}$)
above the $^{56}$Ni core.  ISB effects due to the Coulomb force were taken into account. \replaced{Our results suggest alternatives  to the shape change proposed}{ in  and our results question the ``shape change" conclusions reached} in Ref.~\cite{wimmer2021}. On one hand, the calculated M$_p$(E2) matrix elements, using the DZ effective charges,  appear in line with the experimental data, given the statistical uncertainties. On the other hand, the use of ST effective charges may indicate that $^{70}$Br, rather than $^{70}$Kr,  deviates from the isospin symmetry  expectations and we have proposed a scenario which could explain the Coulomb excitation measurements.  \added{Given the important ISB effects implied by the experimental data,}  perhaps, further experimental work with \replaced{a more sensitive}{an improved} $\gamma$-ray spectrometer (such as a tracking array) should be considered \replaced{to probe the scenarios discussed in this work}{ the before invoking strong ISB effects}. \\
\medskip

\noindent 
{\sl Acknowledgements.~} This material is based upon work supported by the U.S. Department of Energy, Office of Science, Office of Nuclear Physics under Contract  No. DE-AC02-05CH11231 (LBNL).
 AP work is supported in part by the Ministerio de Ciencia, Innovaci\'on y Universidades (Spain), Severo Ochoa Programme SEV-2016-0597 and grant PGC-2018-94583. \\

\end{document}